\documentclass[3p,times,twocolumn]{elsarticle}

\usepackage{amsmath}
\usepackage{ecrc}

\volume{00}

\firstpage{1}

\journalname{Nuclear and Particle Physics Proceedings}

\runauth{A.T. Araudo et al.}

\jid{nppp}

\jnltitlelogo{Nuclear and Particle Physics Proceedings}

\usepackage{amssymb}

\usepackage[figuresright]{rotating}

\begin{document}

\begin{frontmatter}



\dochead{}

\title{Evidence that particle acceleration in hotspots of FR II galaxies is not constrained by synchrotron cooling}

\author[label1,label2]{Anabella T. Araudo}
\address[label1]{Laboratoire Univers et Particules de Montpellier CNRS/Université de Montpellier, Place E. Bataillon, 34095 Montpellier, France}
\address[label2]{University of Oxford, Astrophysics, Keble Road, Oxford OX1 3RH, UK}
\author[label3]{Anthony R. Bell}
\address[label3]{University of Oxford, Clarendon Laboratory, Parks Road, 
Oxford OX1 3PU, UK}
\author[label2]{Katherine M. Blundell}

\begin{abstract}
We study the hotspots of powerful radiogalaxies, where
electrons accelerated at the jet termination shock emit synchrotron radiation.
The turnover of the synchrotron spectrum is typically observed between 
infrared and optical frequencies,
indicating that the maximum energy of non-thermal electrons accelerated
at the shock is $\lesssim$~TeV for a canonical magnetic field of
$\sim$100~$\mu$G. We show that this maximum energy cannot be 
constrained by synchrotron losses as usually assumed, unless the jet density
is unreasonably large  and most of the jet upstream energy goes to 
non-thermal particles. We test this result by considering a sample of hotspots 
observed  at radio, infrared and optical wavelengths.
\end{abstract}

\begin{keyword}
galaxies: active \sep galaxies: jets \sep acceleration of particles \sep radiation mechanisms: non-thermal \sep shock waves  


\end{keyword}

\end{frontmatter}


\section{Introduction}
\label{intro}

\begin{figure}
\includegraphics[width=0.8\columnwidth]{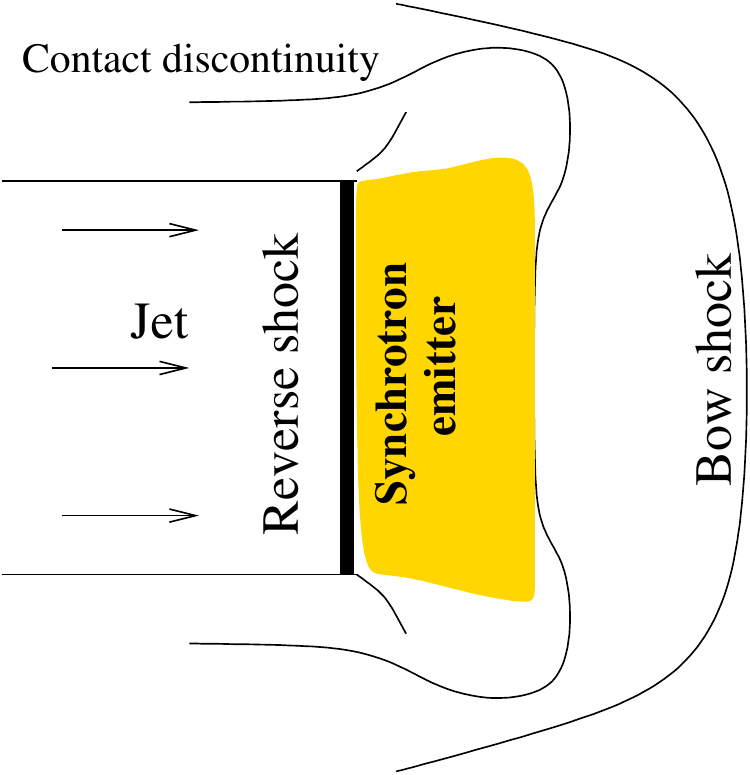}
\caption{Sketch of the standard picture of the jet termination region. 
Particles are accelerated
at the reverse shock, and radiate in the shock downstream region, here
labelled ``Synchrotron emitter''.}
\label{sketch}
\end{figure}

The jet termination region of Fanaroff-Riley (FR) 
radiogalaxies \citep{FR} is characterised by a 
double shock structure separated by a contact discontinuity, as sketched in 
Figure~\ref{sketch}\footnote{Note that the contact discontinuity is
unstable due to the velocity shear and density contrast in both
sides of the discontinuity \cite[e.g.][]{Mizuta_04}.}.
Hotspots are the  downstream region of the 
jet reverse shock, where electrons accelerated by the shock emit
synchrotron radiation. The cut-off of the synchrotron spectrum  
at $\nu_{\rm c} \gtrsim 10^{14}$~Hz typically observed in hotspots  
\cite[e.g.][]{3c273-Natur,Meisenheimer_IR,Tavecchio_05,Stawarz_07,Werner_12}
indicates that the maximum energy of non-thermal  electrons is 
\begin{equation}
\frac{E_{\rm c}}{\rm TeV} \sim 0.2
\left(\frac{\nu_{\rm c}}{10^{14}\,{\rm Hz}}\right)^{\frac{1}{2}}
\left(\frac{B}{100\,\mu{\rm G}}\right)^{-\frac{1}{2}},
\label{Ec}
\end{equation}
where $B$ is the magnetic field \citep{Ginzburg}. 
In some cases, X-rays are also detected and modeled as 
synchrotron self Compton emission
and Compton up-scattering of Cosmic Microwave Background photons 
\cite[e.g.][]{Perlman_10,Wilson_00}. 

Ions can also be accelerated in the jet reverse shock. Given that hadronic losses are very slow in
low density plasmas such as  the termination region of AGN jets, protons might achieve 
energies as large as the limit imposed by the size of the source (i.e. the "Hillas" limit). In particular,  
mildly relativistic reverse shocks with velocity  $v_{\rm sh}\sim c/3$ \cite{Casse_05,Steenbrugge_08} 
might accelerate particles with  Larmor radius
$r_{\rm g} \sim R_{\rm j}$, where $R_{\rm j}$ is the jet width at the
termination region. Particles with such a large $r_{\rm g}$  have energy 
\begin{equation}
\frac{E_{\rm UHECR}}{\rm EeV} \sim 100
\left(\frac{v_{\rm sh}}{c/3}\right)
\left(\frac{B}{100 \,\rm \mu G}\right)
\left(\frac{R_{\rm j}}{\rm kpc}\right)
\label{hillas}
\end{equation}
\citep{Lagage-Cesarsky,Hillas},  and therefore hotspots have been proposed as sources of
Ultra High Energy Cosmic Rays (UHECRs) \citep[e.g.][]{Rachen_93,Norman_95}. 
But, there are two assumptions in Eq.~(\ref{hillas}): 
\begin{enumerate}
\item Particles diffuse in the Bohm regime, i.e. the mean-free path is $\lambda \sim r_{\rm g}$.
\item The magnetic field $B$ persists over distances $\sim R_{\rm j}$ downstream of the shock. 
\end{enumerate}

Damping of the magnetic field in the southern hotspot of the quasar 4C74.26 
was confirmed by modelling the compact synchrotron emission ($\sim$0.1~kpc)  
detected by the MERLIN interferometer at  1.66~GHz \citep{4c7426}.
The very thin synchrotron emitter  would require 
a magnetic field $\sim 2.4$~mG to match the size of the emision region with 
the synchrotron cooling length at 1.66~GHz. 
This value of the magnetic field is about 10 times the upper limit imposed by
the equipartition condition with non-thermal particles (see more details
of the model in \cite{heproV}). Therefore, the compact radio
emission delineates the region within which the magnetic
field is amplified up to
$\sim$100~$\mu$G, and it is damped downstream of the shock. 
The cutoff of the synchrotron spectrum in the southern 
hotspot of 4C74.26 is between  infrared (IR) and optical frequencies, and 
therefore the maximum energy of non-thermal 
electrons accelerated by the termination shock is $E_{\rm c}\sim$0.3~TeV.
Given that the thickness of the MERLIN radio emitter  is larger than the 
synchrotron cooling length of 0.3~TeV-electrons, we interpreted this 
behaviour 
of $E_{\rm c}$ being determined by synchrotron cooling, and then, at
distances $\sim$0.1~kpc downstream of the shock, the magnetic field is damped
as a consequence of the small scale of Weibel turbulence.
However, 0.1~kpc is
much larger than the turbulence decay length predicted by numerical
simulations of Weibel-mediated shocks in plasmas with densities 
$\sim$10$^{-4}$~cm$^{-3}$.

In a recent paper \citep{Araudo_16}, we revisit the assumption of the 
cutoff of the hotspot synchrotron spectrum being determined 
by synchrotron losses. 
Given that the scale-length of magnetic fluctuations
has to be larger than the plasma skin depth $c/\omega_{\rm pi}$,  we show that
$E_{\rm c}$ cannot be determined by synchrotron cooling, as usually assumed, 
unless very extreme conditions are assumed. We consider the sample of hotspots 
observed with high spatial resolution at radio, IR and optical frequencies 
in \cite{Mack_09}. We also show that the Weibel 
instability is not the source of the amplified magnetic field throughout the
whole hotspot emission region since not only does it damp too quickly, 
but also it generates turbulence on a very small scale,
insufficient to accelerate particles up to $E_{\rm c}\sim$TeV for typical 
values of the magnetic field.  We discuss the alternative possibility that the  
turbulence is generated by the Non Resonant Hybrid (NRH) 
instability \citep{Bell_04,Tony_05} which damps less quickly and grows on a 
larger scale. In the present contribution we highlight these results. 

We remark that the results presented  in \cite{Araudo_16} have important implications 
for Eq.~(\ref{hillas}) and
the maximum energy that protons can achieve by being accelerated in 
the jet reverse shock. 
We conclude that  hotspots of FR~II radiogalaxies with IR/optical 
synchrotron cut-off are very poor accelerators of UHECRs.

\section{The reigning paradigm}
\label{hotspots}

The traditional assumption is that $E_{\rm c}$ is determined by synchrotron 
cooling. By equating the synchrotron cooling timescale 
$t_{\rm synchr} \sim 300/(B^2 E_{\rm c})$~s with the acceleration time 
$t_{\rm acc} \sim 20 \mathcal{D}/v_{\rm sh}^2$,
the diffusion coefficient  of electrons with energy  
$E_{\rm c}$ (as in Eq.~\ref{Ec}) is\footnote{The subscript "c,s" in the diffusion coefficient $\mathcal{D}$ 
and in the mean-free path $\lambda$ means that the energy of particles is $E_{\rm c}$ and it is
determined by synchrotron losses (i.e. $t_{\rm synchr}(E_{\rm c})=t_{\rm acc}(E_{\rm c})$).}
\begin{equation}
\frac{\mathcal{D_{\rm c,s}}}{\mathcal{D_{\rm Bohm}}} \sim 10^7
\left(\frac{v_{\rm sh}}{c/3}\right)^2
\left(\frac{\nu_{\rm c}}{10^{14}\,{\rm Hz}}\right)^{-1},
\label{DDBohm}
\end{equation}
\cite[e.g.][]{Stage_06,Kirk_Brian_10}, where 
$\mathcal{D_{\rm Bohm}}\sim r_{\rm g}c/3$ is the Bohm diffusion coefficient.
Note that protons with  energy $\sim E_{\rm c}$ also diffuse with $\mathcal{D_{\rm c,s}}$
and therefore the maximum energy that they can 
achieve is reduced to 10~TeV instead of 100~EeV as expected from the Hillas
constraint in Eq.~(\ref{hillas}).

The  mean-free path of particles diffusing in a medium with $\mathcal{D_{\rm c,s}}$ is
\begin{equation}
\lambda_{\rm c,s} \sim \frac{\mathcal{D}_{\rm c,s}}{c/3} \sim 25
\left(\frac{v_{\rm sh}}{c/3}\right)^{2}
\left(\frac{\nu_{\rm c}}{10^{14}\,{\rm Hz}}\right)^{-\frac{1}{2}}
\left(\frac{B}{100\,\mu{\rm G}}\right)^{-\frac{3}{2}} {\rm pc}.
\label{mfp_c}
\end{equation}
We will show in the next section that $\lambda_{\rm c,s}$ is larger than the maximum value of the mean-free
path imposed by plasma physics.

\section{Revising the  reigning paradigm}
\label{mfa}

The mean-free path of particles in a medium with (small) magnetic-turbulence scale-length $s$ is 
$\lambda \sim r_{\rm g}^2/s$ \citep[e.g.][]{Ostrowski_02,Kirk_Brian_10,lemoine-pelletier-10,Sironi_13}.  
For a hydrogen plasma jet with  proton thermal Lorentz factor
$\sim$1, density  $n_{\rm j}$, and frequency $\omega_{\rm pi}$,  
the ion-skin depth downstream of the shock is
\begin{equation}
\frac{c}{\omega_{\rm pi}} \sim 8.6\times10^8 
\left(\frac{n_{\rm j}}{10^{-4}\,{\rm cm^{-3}}}\right)^{-\frac{1}{2}}\,{\rm cm}.
\label{c_omega_pi}
\end{equation}
Considering that $s$ cannot be smaller than $c/\omega_{\rm pi}$, 
we find the upper-limit $\lambda_{\rm max} = r_{\rm g}^2/(c/\omega_{\rm pi})$. 
In particular,  $\lambda_{\rm max}$ of
the most energetic electrons accelerated at the jet reverse shock is
\begin{eqnarray}\label{lambda_c}
\begin{aligned}
\lambda_{\rm max} &= 
\frac{r_{\rm g,c}^2}{c/\omega_{\rm pi}} \\
&\sim 0.02
\left(\frac{\nu_{\rm c}}{10^{14}\,{\rm Hz}}\right)
\left(\frac{B}{100\,\mu{\rm G}}\right)^{-3}
\left(\frac{n_{\rm j}}{10^{-4}\,{\rm cm^{-3}}}\right)^{\frac{1}{2}}\,{\rm pc},
\end{aligned}
\end{eqnarray}
where 
\begin{equation}
\frac{r_{\rm g,c}}{\rm cm}  \sim 9\times10^{12} 
\left(\frac{\nu_{\rm c}}{10^{14}\,{\rm Hz}}\right)^{0.5}
\left(\frac{B}{100\,\mu{\rm G}}\right)^{-1.5}
\end{equation}
is the Larmor radius of $E_{\rm c}$-electrons ($r_{\rm g,c}\equiv r_{\rm g}(E_{\rm c})$). 
Therefore, the maximum diffusion  coefficient is
\begin{eqnarray}
\begin{aligned}
\frac{\mathcal{D}_{\rm max}}{\mathcal{D}_{\rm Bohm}} = &
\frac{\lambda_{\rm max}}{r_{\rm g,c}}
\sim 3.2\times10^4
\left(\frac{\nu_{\rm c}}{10^{14}\,{\rm Hz}}\right)^{\frac{1}{2}}\\
&\left(\frac{B}{100\,\mu{\rm G}}\right)^{-\frac{3}{2}}
\left(\frac{n_{\rm j}}{10^{-4}\,{\rm cm^{-3}}}\right)^{\frac{1}{2}}.
\label{D_DBohm}
\end{aligned}
\end{eqnarray}

\begin{figure}
\includegraphics[width=0.56\textwidth]{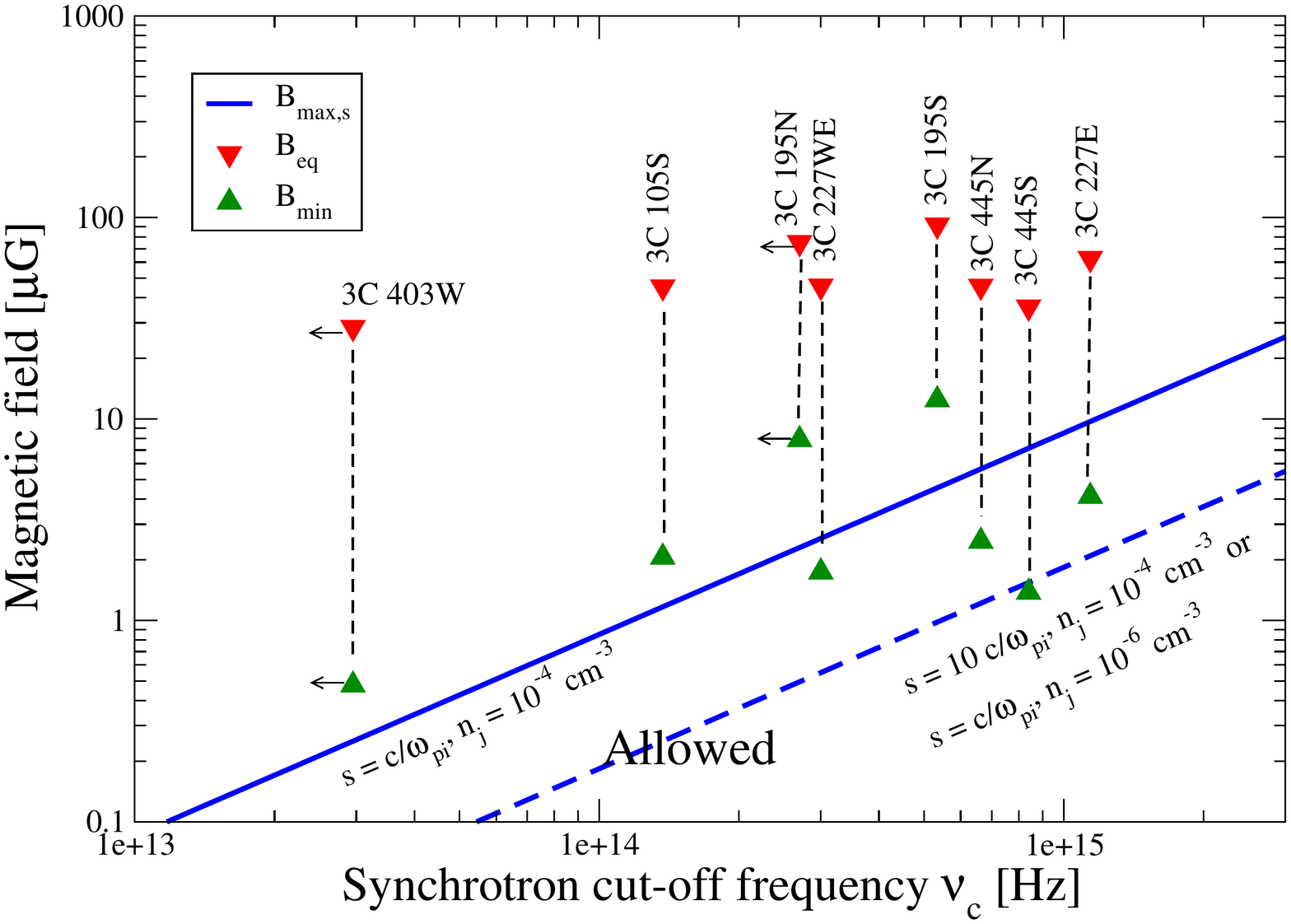}
\caption{Upper limit $B_{\rm max,s}$ for the magnetic field imposed by the 
condition  $\lambda_{\rm c,s} \ge \lambda_{\rm max}$ 
($n_{\rm j}=10^{-4}$~cm$^{-3}$: blue-solid line;
$n_{\rm j}=10^{-6}$~cm$^{-3}$: blue-dashed line). Triangles indicate the
maximum ($B_{\rm eq}$, red triangles) and minimum ($B_{\rm min}$, 
green triangles) field for the sources in Mack et al. (2009).
\label{B_nu_sources}}
\end{figure}

If $E_{\rm c}$ were determined by a competition between shock acceleration 
and synchrotron cooling (the reigning paradigm), 
the mean-free path of $E_{\rm c}$-electrons would be given by Eq.~(\ref{mfp_c}). 
By comparing  $\lambda_{\rm c,s}$ with the upper-limit $\lambda_{\rm max}$,
we find that  
\begin{eqnarray}
\begin{aligned}
\frac{\lambda_{\rm c,s}}{\lambda_{\rm max}} \sim & 3\times10^4
\left(\frac{v_{\rm sh}}{c/3}\right)^2
\left(\frac{\nu_{\rm c}}{10^{14}\,{\rm Hz}}\right)^{-\frac{3}{2}}
\left(\frac{B}{100\,\mu{\rm G}}\right)^{\frac{3}{2}}\\
&\left(\frac{n_{\rm j}}{10^{-4}\,{\rm cm^{-3}}}\right)^{-\frac{1}{2}}.
\label{lambda_ratio}
\end{aligned}
\end{eqnarray}
Setting $\lambda_{\rm c,s} \le \lambda_{\rm max}$ 
(or $\mathcal{D}_{\rm c,s} \le \mathcal{D}_{\rm max}$) implies 
a magnetic field $B\le B_{\rm max,s}$, where
\begin{equation}\label{Bs}
\frac{B_{\rm max,s}}{\rm \mu G} \sim 0.8
\left(\frac{\nu_{\rm c}}{10^{14}\,{\rm Hz}}\right)
\left(\frac{v_{\rm sh}}{c/3}\right)^{-\frac{4}{3}}
\left(\frac{n_{\rm j}}{10^{-4}\,{\rm cm^{-3}}}\right)^{\frac{1}{3}}.
\end{equation}
In Fig.~\ref{B_nu_sources} we plot $B_{\rm max,s}$ for the cases of
$n_{\rm j} = 10^{-4}$ (blue-solid line) and $10^{-6}$~cm$^{-3}$ (blue-dashed line). 
The small values of $B_{\rm max,s}$ 
would require a very large energy density in non-thermal electrons  ($U_e$) in order to 
explain the measured radio synchrotron flux, as described in the next section. 
However, $U_e$ cannot be larger than the jet kinetic energy density 
$U_{\rm kin} = m_pc^3n_{\rm j}(\Gamma_{\rm j}-1)$, where $\Gamma_{\rm j}=1.06$ is
the bulk jet Lorentz factor if the jet velocity is $c/3$:
\begin{equation}\label{Ukin}
\frac{U_{\rm kin}}{\rm erg\,cm^{-3}} \sim 10^{-8}
\left(\frac{\Gamma_{\rm jet} - 1}{0.06}\right)
\left(\frac{n_{\rm j}}{10^{-4}\,{\rm cm^{-3}}}\right).
\end{equation}

\subsection{Synchrotron emission at 8.4~GHz}
\label{synch_vla}

We consider the sample of 8 hotspots observed at 8.4~GHz by \cite{Mack_09}, 
and with  radio-to-optical spectral index $\alpha$ (see Table~\ref{tab_sources}).
Synchrotron emission is produced by non-thermal electrons following a power-law energy distribution 
$N_e=K_eE_e^{-p}$, where  $E_e \ge m_ec^2 \gamma_{\rm min}$ and $p = 2\alpha +1$.
The normalization constant  $K_e$ can be determined from the synchrotron  luminosity at 8.4~GHz 
($L_{8.4}$) emitted in a  volume $V$ \cite[see more details in][]{heproV}, and
$U_e\sim K_e \gamma_{\rm min}^{2-p}/(p-2)$ can be written as 
\begin{eqnarray}\label{Ue}
\begin{aligned}
\frac{U_{e}}{\rm erg\,cm^{-3}}  & \sim  10^{-9}
\left(\frac{p-2}{0.5}\right)^{-1}  
\left(\frac{\gamma_{\rm min}}{100}\right)^{2-p}\\
&\left(\frac{L_{8.4}}{10^{41}\,{\rm erg\,s^{-1}}}\right)
\left(\frac{V}{{\rm kpc^{3}}}\right)^{-1}
\left(\frac{B}{100\rm \mu G}\right)^{\frac{-p-1}{2}}.
\end{aligned}
\end{eqnarray}

By setting the extreme condition $U_e = U_{\rm kin}$ we find a lower limit in the magnetic field 
($B_{\rm min}$) needed to explain the synchrotron emission at 8.4~GHz
\begin{equation}\label{Bmin}
\begin{aligned}
\frac{B_{\rm min}}{\rm \mu G}  & \sim  27^{\frac{3.5}{p+1}}
\left(\frac{\gamma_{\rm min}}{100}\right)^{\frac{4-2p}{(p+1)}}
\left(\frac{L_{8.4}}{10^{41}\,{\rm erg\,s^{-1}}}\right)^{\frac{2}{p+1}}\\
&\left[\left(\frac{\Gamma_{\rm j} - 1}{0.06}\right)
\left(\frac{p-2}{0.5}\right)
\left(\frac{V}{{\rm kpc^{3}}}\right)
\left(\frac{n_{\rm j}}{10^{-4}\rm cm^{-3}}\right)\right]^{\frac{-2}{p+1}}.
\end{aligned}
\end{equation}
We compute $B_{\rm min}$ for all the sources in \cite{Mack_09}; see
Table~\ref{tab_sources} and Fig.~\ref{B_nu_sources} (green-triangles up) where we also list and plot
the value of the magnetic field $B_{\rm eq}$ in equipartition with non-thermal electrons 
(i.e. $B_{\rm eq}^2/(8\pi)=U_e$).
We can see that $B_{\rm min} > B_{\rm max,s}$ (blue-solid line)
for  sources 3C\,105S, 3C\,195N, 3C\,195S and 3C\,403W whereas $B_{\rm min} < B_{\rm max,s}$
for 3C\,227WE, 3C\,227E, 3C\,445N and 3C\,445S. Note however that:
\begin{itemize}
\item The jet density $n_{\rm j} \sim 10^{-4}$~cm$^{-3}$ is the upper limit found in Cygnus~A and
3C475, and therefore we expect values of  $B_{\rm min}$ greater than those 
plotted in Fig.~\ref{B_nu_sources} when the jet density is smaller than
$10^{-4}$~cm$^{-3}$ ($B_{\rm min} \propto n_{\rm j}^{-(p+1)/2}$). 
On the other hand, $B_{\rm max,s} \propto n_{\rm j}^{1/3}$ and therefore $B_{\rm max,s}$
decreases when smaller values of $n_{\rm j}$ are considered and
the ratio $B_{\rm min}/B_{\rm max,s} \propto n_{\rm j}^{-(p+5/6)}$. In particular,
the blue-dashed line in Fig.~\ref{B_nu_sources} corresponds to the case of
$n_{\rm j} = 10^{-6}$~cm$^{-3}$ and $s = c/\omega_{\rm pi}$.   
In such a case,  sources 
3C\,227WE, 3C\,227E, 3C\,445N and 3C\,445S move to the regime where
$B_{\rm min} > B_{\rm max,s}$. The minimum value of the jet density required to
match $B_{\rm min} = B_{\rm max,s}$ is 
listed in Table~\ref{tab_sources} for all the sources
considered in this paper. We can see for instance that the source 3C~195N 
necessitates $n_{\rm j} > 6.5\times10^{-4}$~cm$^{-3}$ to satisfy the 
conditions $\lambda_{\rm c,s}<\lambda_{\rm max}$ (i.e. $s>c/\omega_{\rm pi}$) and 
$U_e < U_{\rm kin}$.

\item Even when jets in FR galaxies are expected to be  
perpendicular to the line of sight, small departure from the plane of the 
sky (i.e. $\theta_{\rm j}<90^{\circ}$, where $\theta_{\rm j}$ is the angle between 
the jet and the line of sight) reduces the  size of the shock downstream 
region.  In such a case, $B_{\rm min} \propto V^{-2/(p+1)}$ 
increases whereas $B_{\rm max,s}$ remains constant. 
Therefore, the situation is even more strongly ruled out when 
$\theta_{\rm j}<90^{\circ}$.
\end{itemize}

Even in the case that the extreme conditions
discussed before are assumed, the large value of the diffusion coefficient
required to be $E_{\rm c}$ determined by synchrotron losses
($\mathcal{D}_{\rm c,s}/\mathcal{D}_{\rm Bohm}\sim 10^6$-$10^7$, see 
Eq.~(\ref{DDBohm}))
cannot be explained in any well-established theoretical framework \cite[see Section~4 in][]{Araudo_16}.
We therefore suggest that the IR/optical cutoff in the synchrotron spectrum has a different cause.

\begin{table*}
\caption{Physical parameters of the sources considered in this paper.
The synchrotron cut-off frequency ($\nu_{\rm c}$) and spectral index ($\alpha$) are taken from Mack et al. (2009);  and $p=2\alpha+1$.
The synchrotron luminosity at 8.4~GHz is calculated as 
$L_{8.4} = S_{8.4} 8.4\times10^9$, where 
$S_{8.4}$ is the measured flux. The hotspot volume $V$  is 
calculated from the angular sizes 
tabulated in Table~5 in Mack et al. (2009) together with 
$S_{8.4}$. The magnetic field in equipartition with non thermal electrons ($B_{\rm eq}$) is calculated in 
Araudo et al. (2016) and it is similar to the values given in Mack et al. (2009). $B_{\rm min}$ is the magnetic field required to emit $L_{8.4}$ when $U_e = U_{\rm kin}$, and $B_{\rm max,s}$ is the 
magnetic field that satisfies the condition $\lambda_{\rm c,s}=\lambda_{\rm max}$. }
\centering
\label{tab_sources}
\begin{tabular}{lcc|ccccccc}
\hline\hline
Source&$\nu_{\rm c}$&$\alpha$& $p$& $L_{8.4}$&$V$ &$B_{\rm eq}$& $B_{\rm min}$ &$B_{\rm max,s}$ & $n_{\rm j,min}$\\
& [$10^{14}$~Hz]&&&[erg/s]&[kpc$^3$] &[$\mu$G]&[$\mu$G]&[$\mu$G] &[cm$^{-3}$]\\
\hline
3C\,105S &1.37  &0.75 &2.5 &1.42$\times$10$^{42}$& 1205.63& 45.27& 
2.06&1.16&1.92$\times$10$^{-4}$\\ 
3C\,195N & $<$2.70&0.95 &2.9 &1.15$\times$10$^{41}$& 38.12& 75.11& 7.89&
2.30&6.51$\times$10$^{-4}$\\
3C\,195S & 5.34  &1.00 &3.0 &1.71$\times$10$^{41}$& 33.58& 91.76&  12.45&
4.55&3.42$\times$10$^{-4}$\\
3C\,227WE& 3.00  &0.65 &2.3 &3.19$\times$10$^{40}$& 19.26& 45.63&  1.74&
2.55&6.78$\times$10$^{-5}$\\
3C\,227E &11.4  &0.75 &2.5 &7.14$\times$10$^{40}$& 17.99& 62.60 &  4.12&
9.71&3.96$\times$10$^{-5}$\\
3C\,403W & $<$0.29&0.55&2.1 &3.95$\times$10$^{40}$& 167.9& 28.46 &  0.48&
0.25&1.96$\times$10$^{-4}$\\
3C\,445N & 6.63  &0.85 &2.7 &2.18$\times$10$^{40}$& 29.36& 45.60&  2.47&
5.65&3.97$\times$10$^{-5}$\\
3C\,445S & 8.40  &0.80 &2.6 &5.04$\times$10$^{40}$& 139.42& 35.94 &  1.38&
7.15&1.60$\times$10$^{-5}$\\
\hline\hline
\end{tabular}
\end{table*}

\section{Particle acceleration and magnetic field amplification  in quasi-perpendicular shocks}
\label{perp}

If synchrotron losses are not relevant, the maximum energy is ultimately
determined by the ability to scatter particles downstream of the shock.
We consider that the amplified hotspot magnetic field $B$ is turbulent, 
and that the large scale background field downstream of the reverse shock 
($B_{\rm jd}$) is nearly perpendicular to the shock normal. 

\subsection{Diffusive shock acceleration  in quasi-perpendicular shocks}
\label{DSA_perp}

To accelerate particles  up to an energy 
$E_{\rm c}$ via a diffusive mechanism in a perpendicular shock,  the mean-free path 
$\lambda_{\rm c}\sim r_{\rm g,c}(B)^2/s$
in the shock downstream region 
has to be smaller than Larmor radius in $B_{\rm jd}$ 
\citep[e.g.][]{Kirk_Brian_10, lemoine-pelletier-10,Brian_14}. Otherwise, the particles are carried away downstream 
by the magnetic field and acceleration is halted.  
The condition $\lambda_{\rm c} \le r_{\rm g,c}(B_{\rm jd})$
is satisfied when the magnetic-turbulence scale-length is $s \ge s_{\perp}$, where
\begin{equation}\label{s_nrh}
\begin{aligned}
s_{\perp}&\equiv\frac{E_{\rm c}}{eB}\left(\frac{B_{\rm jd}}{B}\right) \\
&= 5\times10^{11}
\left(\frac{\nu_{\rm c}}{10^{14}\,{\rm Hz}}\right)^{\frac{1}{2}}
\left(\frac{B_{\rm jd}}{\rm \mu G}\right)
\left(\frac{B}{\rm 100\mu G}\right)^{-\frac{5}{2}}\,{\rm cm},
\end{aligned}
\end{equation}
and $s_{\perp}$ is the Larmor radius of protons (and electrons) with energy
\begin{equation}\label{gamma_nrh}
\begin{aligned}
E_{\rm s_{\perp}} &= E_{\rm c}\left(\frac{B_{\rm jd}}{B}\right) =
0.07 E_{\rm c}
\left(\frac{B_{\rm j}}{\rm \mu G}\right)
\left(\frac{B}{100 \rm \mu G}\right)^{-1}\\
&\sim10\left(\frac{\nu_{\rm c}}{10^{14}\,{\rm Hz}}\right)^{\frac{1}{2}}
\left(\frac{B_{\rm jd}}{\rm \mu G}\right)
\left(\frac{B}{\rm 100\mu G}\right)^{-\frac{5}{2}}\,\,{\rm GeV}
\end{aligned}
\end{equation}
in a magnetic field $B$ (i.e. $s_{\perp} = E_{\rm s_{\perp}}/qB$). Note that  $s_{\perp}$ in Eq.~(\ref{s_nrh})
is greater than $c/\omega_{\rm pi}$ in Eq.~(\ref{c_omega_pi}), as 
required. However,  $s > s_{\perp} \sim$500$c/\omega_{\rm pi}$
(for typical values considered in this paper) cannot be fulfilled by 
Weibel-generated turbulence with scale $\sim c/\omega_{\rm pi}$.

\subsection{Non-resonant hybrid instabilities in quasi-perpendicular shocks}
\label{hotspots_mfd}

Turbulence on a scale greater than $c/\omega_{\rm pi}$ may be excited through
the NRH instability \citep{Bell_04,Tony_05}.
In the simplest form of it, the cosmic ray (CR) Larmor radius 
in the unperturbed background field is much greater than the 
wavelength of field perturbations  and therefore the streaming
of CRs carrying the electric current $j_{\rm cr}$
is undeflected. The force -$\vec j_{\rm cr}\times \vec B$
acts to expand loops in the magnetic field, and therefore $B$ 
increases. This produces an increment in 
-$\vec j_{\rm cr}\times \vec B$ and generates a positive feedback loop 
that drives the NRH instability and amplifies the magnetic field.
For the diamagnetic drift in the plane of the shock to amplify
the magnetic field,
the NRH growth rate has to be sufficient for the instability 
to grow through $\sim$10 e-foldings at the maximum growth rate 
$\Gamma_{\rm max}$ \citep{Bell_04,Tony_05,Bell_rev_14} in the time 
$t_{\perp}\sim r_{\rm g}(B_{\rm js})/v_{\rm d}$ the plasma 
flows through a distance $r_{\rm g}(B_{\rm js})$ in the downstream region at velocity 
$v_{\rm d}\sim v_{\rm sh}/4$,
where $r_{\rm g}(B_{\rm js})$ 
is the Larmor radius in the ordered field $B_{\rm js}$. That is, the condition
$\Gamma_{\rm max} t_{\perp} > 10$ must be satisfied. 
If the field is strongly amplified, the instability can be expected
to saturate when its characteristic scale grows to the Larmor radius of the
CR driving the instability. Thus, $s_{\perp}$ in Eq.~(\ref{s_nrh}) can be expected to
match the Larmor radius of the highest energy CR driving the instability.
If these CR have an energy $E_{\rm nrh}$, then  $E_{\rm nrh}\sim E_{\rm s_{\perp}}$.

In order to check that there is enough energy in $E_{\rm nrh}$-protons
to excite the non-resonant turbulence\footnote{Note that $E_{\rm s_{\perp}}\sim 100\,m_pv_{\rm sh}^2$ 
if $v_{\rm sh} \sim c/3$ and therefore $E_{\rm nrh}$-CRs are mildly supra-thermal protons.}, we consider whether the number of e-foldings required to amplify the magnetic field up to the saturation value 
is of the order of 10 \citep{Bell_04, Tony_05}. The condition  $\Gamma_{\rm max} t_{\perp}> 10$ leads to 
\begin{equation}
\begin{aligned}
\eta \gtrsim 0.04 \left(\frac{B_{\rm j}}{\mu\rm G}\right)
\left(\frac{\Gamma_{\rm j} - 1}{0.06}\right)^{-\frac{1}{2}} 
\left(\frac{n_{\rm j}}{10^{-4}\,{\rm cm^{-3}}}\right)^{-\frac{1}{2}},
\end{aligned}
\end{equation}
where $\eta$ is the acceleration efficiency. Given that 
particles accelerated in relativistic shocks follow a power-law energy
distribution  steeper than the canonical distribution (i.e. $p>2$ as is clear in Table~\ref{tab_sources}), 
the CR pressure  is dominated by low energy particles.  
Therefore, the condition for NRH instability growth is that the acceleration 
efficiency of low energy CR has to be
$\eta \sim 0.04$ for characteristic values considered in this paper. 
Such a value of $\eta$ is very reasonable. 

From these  estimations we can conclude that NRH instabilities 
generated by CRs with energies $\sim E_{\rm nrh}$ can grow fast enough 
to amplify the jet magnetic field 
from $\sim$1 to 100~$\mu$G and accelerate particles up to energies
$E_{\rm c} \sim 0.2$~TeV observed in the hotspots of FR~II radiogalaxies.
The advantage of magnetic turbulence being generated by CRs current is that
the amplified magnetic field persists over long distances downstream of the 
shock, and therefore particles accelerated very near the shock can emit
synchrotron radiation far downstream.

\section{Summary and conclusions}
\label{disc}

We have investigated the physical mechanism that constraints the
maximum energy of particles accelerated at the jet reverse shock in FR~II radiogalaxies. 

By equating the acceleration and synchrotron cooling timescales,  
the mean free path $\lambda_{\rm c,s}$ of $E_{\rm c}$-electrons 
is greater than the maximum value $\lambda_{\rm max} = r_{\rm g,c}^2/(c/\omega_{\rm pi})$
for reasonable values of the magnetic field and jet density (see Eq.~(\ref{lambda_ratio})).
By  considering a sample of 8 hotspots observed 
at optical, IR and radio wavelengths \citep{Mack_09}, we show that 
unreasonable large values of the jet density would be required 
(see $n_{\rm j,min}$ in Table~\ref{tab_sources}) to explain the synchrotron flux at 8.4~GHz
if $E_{\rm c}$  were determined  by synchrotron cooling (see Fig.~\ref{B_nu_sources}). 

If synchrotron losses are not relevant, the maximum
energy is ultimately determined by the ability to scatter particles 
downstream of the shock. 
Weibel-mediated shocks generate the magnetic field 
and accelerate particles \cite[e.g.][]{Spitkovsky_08II,Martins_09}. 
However, the characteristic scale of Weibel turbulence
cannot account for the cut-off of the synchrotron spectrum observed in 
hotspots, nor  the large extension of the hotspot synchrotron emission, 
much larger than the magnetic decay of $\sim 100c/\omega_{\rm pi}$ predicted
by numerical simulations.
A viable alternative is that
turbulence is generated by the streaming of CRs with energy 
$E_{\rm nrh}\sim E_{\rm c}B_{\rm jd}/B \sim 0.01 E_{\rm c}$ 
(see Sect.~\ref{perp}). The amplified magnetic 
field has a scale-length of the order of the Larmor radius of 
$E_{\rm nrh}$-protons and persists over long distances downstream of the 
shock, accounting for the extension of the synchrotron emitter. 

In a work in progress, we apply our arguments
to the  very well known source Cygnus~A 
for which well resolved and multi-wavelength data are  
available \cite{Pyrzas_15}. We will show 
that the primary hotspot in the western jet of Cygnus~A is a clear case where 
the maximum energy of electrons accelerated in the jet reverse shock is not constrained by 
synchrotron cooling.

\section*{Acknowledgements}

A.T.A. thanks the organisers of the conference for their kind hospitality. 
The research leading to this article has received funding
from the European Research Council under the European
Community's Seventh Framework Programme (FP7/2007-2013)/ERC grant agreement 
no. 247039. We acknowledge support from the UK
Science and Technology Facilities Council under grant No. ST/K00106X/1.

\bibliographystyle{elsarticle-num}
\bibliography{biblio_CRs}

\end{document}